\documentclass[
english,
aps,
prl,
twocolumn,
showpacs,superscriptaddress,groupedaddress,footinbib,reprint
]{revtex4}
\usepackage{graphicx}
\usepackage{amsmath}
\usepackage{comment}
\usepackage{subfigure}
\usepackage{color}
\usepackage{soul}


\let \ShowFixme = 1
\newif\ifShowFixme
        \ifx\ShowFixme\undefined
        \ShowFixmefalse
\else
        \ShowFixmetrue
\fi
\ifShowFixme
        \newcommand{\fixme}[1]{{{\bf(FIXME: #1)}}}
\else
        \newcommand{\fixme}[1]{}
\fi

\begin{document}

\title{Dynamics of Ion Transport in Ionic Liquids}

\author{Alpha A Lee}
\affiliation{Mathematical Institute, Andrew Wiles Building, University of Oxford, Woodstock Road, Oxford OX2 6GG, United Kingdom}

\author{Svyatoslav Kondrat}
\affiliation{IBG-1: Biotechnology, Forschungszentrum J\"{u}lich, 52425 J\"{u}lich, Germany}
\affiliation{Department of Chemistry, Faculty of Natural Sciences, Imperial College London, SW7 2AZ, UK}

\author{Dominic Vella}
\affiliation{Mathematical Institute, Andrew Wiles Building, University of Oxford, Woodstock Road, Oxford OX2 6GG, United Kingdom}

\author{Alain Goriely}
\affiliation{Mathematical Institute, Andrew Wiles Building, University of Oxford, Woodstock Road, Oxford OX2 6GG, United Kingdom}

\begin{abstract}

A gap in understanding the link between continuum theories of ion transport in ionic liquids and the underlying microscopic dynamics has hindered the development of frameworks for transport phenomena in these concentrated electrolytes. Here, we construct a continuum theory for ion transport in ionic liquids by coarse graining a simple exclusion process of interacting particles on a lattice. The resulting dynamical equations can be written as a gradient flow with a mobility matrix that vanishes at high densities. This form of the mobility matrix gives rise to a charging behaviour that is different to the one known for electrolytic solutions, but which agrees qualitatively with the phenomenology observed in experiments and simulations. 

\end{abstract}

\makeatother
\maketitle


Room temperature ionic liquids play an increasingly important role as electrolytes in electrochemical and electromechanical applications ranging from actuators \cite{nemat2008electrochemomechanics,liu2010influence, colby2013electroactuation} to supercapacitors \cite{gogotsi:sci:06, largeot2008relation, kondrat:jpcm:11, simon_gogotsi:acr:13, kondrat2014single}. Ionic liquids differ from traditional electrolytes in that they consist only of positive and negative ions without any solvent. Recent theoretical calculations \cite{Lee2014} suggest that ionic liquids are concentrated electrolytes, and should not be modelled as a weak electrolyte (with solvent consisting of ion-pairs), as suggested elsewhere \cite{gebbie2013ionic}.

In many important technological applications, ionic liquids are close to an electrified interface \cite{fedorov2014ionic}. Although the equilibrium structure of the electrical double layer is relatively well studied, understanding the dynamic response of ionic liquids to an applied potential or surface charge is more challenging because of the difficulty in identifying an appropriate non-equilibrium dynamic framework. 
Previous theoretical studies \cite{kilic2007steric,zhao2011diffuse,jiang2014time, jiang2014kinetic,yochelis2014transition,yochelis2014spatial} relied on the dynamical density functional theory \cite{marconi1999dynamic,marconi2000dynamic}, in which the ion flux, $\mathbf{j}_{\pm}$, is related to the ion density fields $c_{\pm}$ and free energy density functional $F[c_\pm]$ via
\begin{equation}
\mathbf{j}_\pm = - M_\pm c_\pm \nabla \left( \frac{\delta F}{\delta c_\pm} \right),
\label{ddft}
\end{equation}
where $M_\pm$ is the cation/anion mobility. A key assumption in the derivation of (\ref{ddft}) is that ion density is low compared to an underlying solvent bath \cite{espanol2009derivation}. This assumption may become problematic, as can be illustrated by substituting the lattice gas free energy, $\mathcal{F}_{ex} = (k_B T/a^3) \int_\Omega [a^3 c \log (a^3 c) + (1-a^3 c) \log(1-a^3 c)] \; \mathrm{d}^3x$ where $\Omega$ is system's volume, into (\ref{ddft}); applying the continuity equation, we obtain
\begin{equation}
\frac{\partial c}{\partial t} = D \nabla \cdot \left( \frac{\nabla c}{1-a^3 c}\right),
\label{ddft_HS}
\end{equation}
where $D = M/k_B T$ is the diffusion constant, $k_B$ the Boltzmann constant and $T$ temperature. However, the continuum limit of a system of particles on a lattice with lattice constant $a$ undergoing a simple exclusion process is well known, and leads to the linear diffusion equation $\partial c / \partial t = D \nabla^2 c$ for the particle density $c$ \cite{spohn1991large}, rather than~(\ref{ddft_HS}).


\begin{figure}
\includegraphics[scale=0.5]{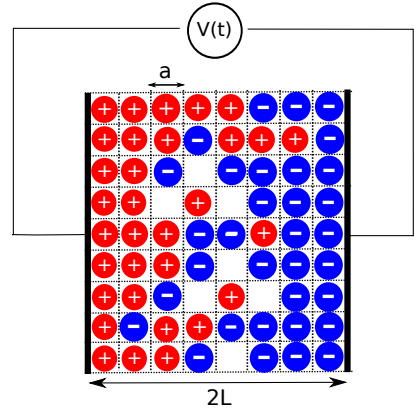}
\caption{Schematic of the system under consideration: cations and anions on a lattice of lattice constant $a$.}
\label{schematic}
\end{figure}


Lattice gas models of ions, first proposed by Bikerman \cite{bikerman1942xxxix} in the 1940s, are commonly used as simple models of ionic liquids~\cite{kilic2007steric,kornyshev2007double,bazant2009towards,bazant2011double,zhao2011diffuse} in equilibrium. The goal of this paper is to derive a consistent model for the dynamics of ions in solvent-free ionic liquids, and analyse the dynamics of electrical double layer formation. We map the system onto a lattice and take the continuum limit of the microscopic reference kinetics of a discrete symmetric exclusion process (see Figure \ref{schematic}).  This reference kinetics is a natural one to consider as ion motion in a concentrated assembly is physically akin to particles ``hopping''  on a lattice \footnote{Simulations show that the nanostructure of ionic liquids with long alkyl chains can be considered as a microphase-separated network of polar and non-polar domains \cite{canongia2006nanostructural,santos2007ionic}. It is thus reasonable to assume that ions move by jumping between polar domains.}. Similar reference kinetics were successfully used to model spinodal decomposition in alloys \cite{gouyet1993atomic,plapp1997surface,plapp1999spinodal,gouyet2003description,petrishcheva2012exsolution}, and were shown to be a microscopic basis for the Cahn-Hilliard equation \cite{giacomin1996exact,giacomin1997phase,giacomin1998phase}. Our approach has the advantage that steric exclusion is accounted for at the level of dynamics.





We first consider a one-dimensional lattice of lattice constant $a$ (corresponding to ion diameter $a$ in the continuum limit) for simplicity, and later generalize our results to higher dimensions. We consider a discrete-time dynamics in which particles can only move between nearest-neighbour lattice sites between time $t$ and $t + \Delta t$. Denoting by $S^{\alpha}_i(t)\in \{0,1\}$ the occupancy of the $i^{\mathrm{th}}$ lattice site at time $t$ by ion of type $\alpha = \{ +,- \}$, the evolution master equation for $S^{\alpha}_i$ reads 
\begin{multline}
S^{\alpha}_i(t+\Delta t) = r^{\alpha}_{i \to i+1} S^{\alpha}_i S_{i+1} 
	+ r^{\alpha}_{i+1\to i}  S^{\alpha}_{i+1} (1 - S_i) \\ 
	+ r^{\alpha}_{i \to i-1}  S^{\alpha}_i S_{i-1} 
	+ r^{\alpha}_{i-1\to i} S^{\alpha}_{i-1} (1 - S_i)  \\
	+ \left(1 - r^{\alpha}_{i\to i+1} - r^{\alpha}_{i\to i-1} \right)S_i^\alpha ,
\label{kinetic_eq}
\end{multline} 
where the $S_i^{\alpha}$ on the right hand side are taken at time $t$ and $S_i = S_i^+ + S_i^-$; $r_{i\to i \pm 1}^\alpha$ is unity if particle $\alpha$ at site $i$ attempts to jump to site $i \pm 1$ and is zero otherwise: this transition propensity takes into account the long-ranged interparticle interactions as explained below. The first term of Eq.~(\ref{kinetic_eq}) ensures that there will be particle $\alpha$ at site $i$ and time $t + \Delta t$, if it is there at time $t$, attempts to move to site $i+1$ and finds that site occupied. The second term describes a possible transition from site $i+1$ to site $i$ when there is no particle at site $i$. The third and forth terms describe the same processes between sites $i$ and $i-1$. Finally, the last term corresponds to a particle $\alpha$ at site $i$ that does not attempt to leave it during the interval $\Delta t$. 



An ensemble average of $r_{i\to j}^\alpha$ with $|i-j|=1$ gives a transition rate that satisfies the detailed balance and can thus be related to the Boltzmann factor by
\begin{align} 
\label{eq:r_av}
\langle r^{\alpha}_{i \to j} \rangle  = \frac{1}{2} e^{-(V^{\alpha}_{j} - V^{\alpha}_i)/2k_BT},
\end{align} 
where
\begin{align} 
\label{eq:U}
V^{\alpha}_i = \sum_{j \neq i}\sum_\beta U_{\alpha\beta}(|i-j|) S^{\beta}_j
\end{align} 
is a potential acting on particle $\alpha$ at site $i$ due to all remaining particles, and $U_{\alpha\beta}(|i-j|)$ is the microscopic (electrostatic) interaction potential between the particles. In the absence of long-ranged interactions ($U_{\alpha\beta} = 0$), there is no preferred direction of motion, thus $\langle r^{\alpha}_{i \to i\pm 1} \rangle = 1/2$.





The continuum evolution equation can be obtained by rescaling the lattice indices by the lattice spacing $a$, and introducing the minimal lattice volume $v$ ($v = a$ in 1-D) as well as the ensemble average concentrations $c_{\alpha}(x=ai) = v^{-1}\langle S^\alpha_i\rangle$ and mean potentials $\mu_\alpha^{l}(x) = \sum_\beta \int_\Omega U_{\alpha\beta}(|x-x'|) c_\beta(x') dx'$. Taking the average of both sides of Eq.~(\ref{kinetic_eq}), and applying the mean field approximation, $\langle S^{\alpha}_i S^{\beta}_j \cdots\rangle \approx \langle S^{\alpha}_i\rangle \langle S^{\beta}_j\rangle \cdots$, we can expand the resulting expression in a power series in $a$ and $\Delta t$ to obtain
\begin{multline}
\label{eq:kin1}
\frac{1}{D} \frac{\partial c_\alpha}{\partial t} = vc_\alpha \frac{ \partial^2 c}{\partial x^2} 
	+ (1-vc) \frac{\partial^2 c_\alpha}{\partial x^2} 
	+ c_\alpha(1-v c) \frac{1}{k_BT}\frac{ \partial ^2 \mu_\alpha^l}{\partial x^2}  \\
	- \left[v c_\alpha \frac{\partial c}{\partial x}- (1-v c) \frac{ \partial c_\alpha}{\partial x} \right] 
		\frac{1}{k_B T} \frac{ \partial \mu_\alpha^l}{\partial x},
\end{multline}
where we have defined $c = c_+ + c_-$ as the total ion density, and identified $D = a^2/(2 \Delta t)$ as the self-diffusion coefficient. 



To generalize Eq.~(\ref{eq:kin1}) to higher dimensions, we assume that the fluxes along different axes are decoupled. Introducing the mobility matrix
\begin{align}
	\label{eq:M}
	\mathcal{M} =  \frac{D}{k_B T}
	\left( 
	\begin{array}{cc}
		 c_+ (1-vc) & 0 \\
	        0 &  c_- (1- vc)  
	\end{array}
	\right), 
\end{align} the higher-dimensional version of \eqref{eq:kin1} is
\begin{align}
	\frac{\partial \boldsymbol{c}}{\partial t} = \nabla \cdot (\mathcal{M} \nabla \boldsymbol{\mu}),
\label{lattice_eq_main}
\end{align}
where $\boldsymbol{c}^T = (c_+, c_-)$, $\boldsymbol{\mu}^T = (\mu_+, \mu_-)$, and $\mu_\pm$ is defined by $\mu_\pm = \delta \mathcal{F} / \delta c_\pm$ where
\begin{multline}
\label{free_energy}
	\mathcal{F}[c_\pm] 
	= \frac{1}{2} \sum_{\alpha\beta} \int_\Omega 
	c_\alpha(\mathbf{x}_1) U_{\alpha\beta} (|\mathbf{x}_1 - \mathbf{x}_2|) c_\beta(\mathbf{x}_2) 
		\; \mathrm{d}\mathbf{x}_1 \mathrm{d} \mathbf{x}_2 \\ 
	 + \frac{k_B T}{v} \int_\Omega \left[\sum_\alpha v c_\alpha \log(vc_\alpha)
		+ (1-vc) \ln(1-v c) \right] \mathrm{d}\mathbf{x}.
\end{multline}

Equation (\ref{lattice_eq_main}) is the continuum kinetic equation for an interacting two component system. Note the important physical constraint that the evolution equation for the total concentration, $c$, in the absence of long-ranged interactions ($U_{\alpha \beta}=0$), reduces to the linear diffusion equation. This constraint, as explained above, respects the fact that the underlying dynamics of our reference system is a simple exclusion process on a lattice. Continuum kinetic equations with the same mobility function as (\ref{lattice_eq_main}) have been proposed in the literature in the context of the modified Cahn-Hilliard equation \cite{nauman2001nonlinear}, and phase-field models of Li-ion batteries \cite{ferguson2012nonequilibrium,zeng2014phase}. 

To apply Equation (\ref{lattice_eq_main}) to an ionic liquid system, we introduce a characteristic length scale $l_c$ for short-ranged interactions, and split the Coulomb potential $U_{\alpha \beta} = U^{sr}_{\alpha \beta} + U^{lr}_{\alpha \beta}$, where  $U_{\alpha \beta}^{sr}(x) = q_\alpha q_\beta l_B e^{-x/l_c}/x$, $U_{\alpha \beta}^{lr}(x)  = q_\alpha q_\beta l_B (1-e^{-x/l_c})/x$ \cite{chen2006local,santangelo2006computing}, and $l_B = e^2/(4 \pi \epsilon_0 \epsilon k_B T)$ is the Bjerrum length. Below the length scale $l_c$, it is actually the hard core exclusion that matters rather than Coulomb interaction, thus $U^{sr}_{\alpha\beta}$ can be neglected. This truncation of the Coulomb potential is necessary as our mean-field approach underestimates steric correlations, and as such the divergence of the Coulomb interaction at the origin renders electrostatic interactions effectively too strong. We thus write
\begin{equation}
\frac{U_{\alpha\beta}(x)}{k_B T} 
	\approx \frac{U^{lr}_{\alpha \beta}(x)}{k_BT}
	= q_\alpha q_\beta l_B \frac{1-e^{-x/l_c}}{x}. 
\label{Coulomb_eff}
\end{equation}
This decomposition of the Coulomb potential is not unique --- the exponential function is chosen phenomenologically and for mathematical convenience.

Introducing the local electric field $u$, and exploiting the Green function, one can rewrite the non-local integro-differential equation \eqref{lattice_eq_main} as a set of coupled partial differential equations
\begin{align}
&(1 - l_c^2 \nabla^2)\nabla^2 u = - 4 \pi l_B (c_+ - c_-),   \label{AKK_overscreening} \\
&  \frac{\partial c_{\pm}}{\partial t} = D \nabla \cdot c_{\pm} (1- v c) \nabla  \left[ \pm u + \ln \left( \frac{v c_\pm}{1- v c}\right)\right]. \label{mod_PNP}
\end{align}
Equation (\ref{AKK_overscreening}) is identical to the modified Poisson equation derived phenomenologically in \cite{bazant2011double} using a gradient expansion of a nonlocal electrostatic kernel. We note that \cite{bazant2013theory}  took the variational approach of \cite{bazant2011double} to develop a framework for charge-transfer reaction kinetics, with the resulting equation similar to (\ref{mod_PNP}). Here we provided a microscopic statistical derivation of the kinetics of ion transport.  

\begin{figure}[!ht]
\subfigure[]{\includegraphics[scale=0.20]{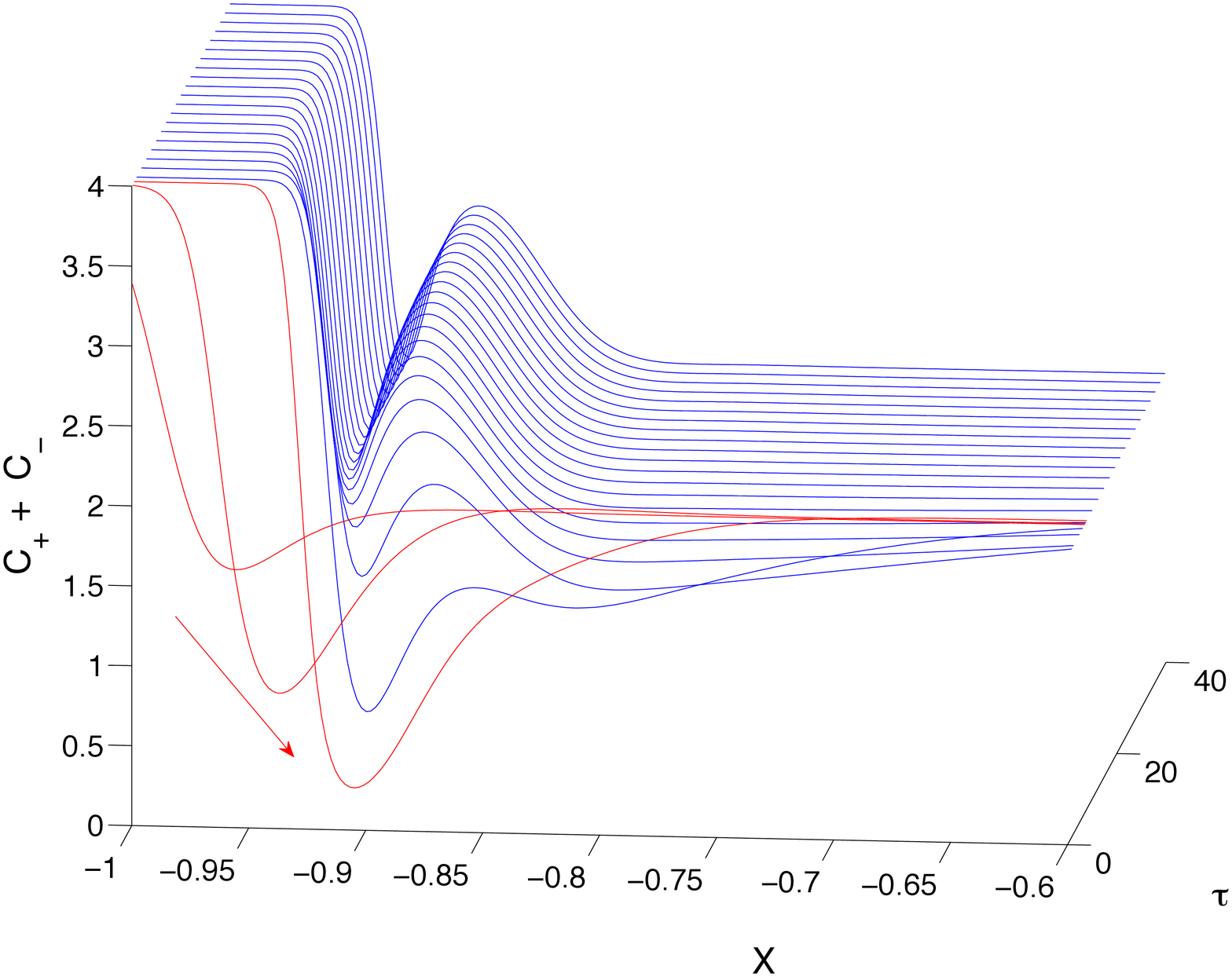}} 
\subfigure[]{\includegraphics[scale=0.20]{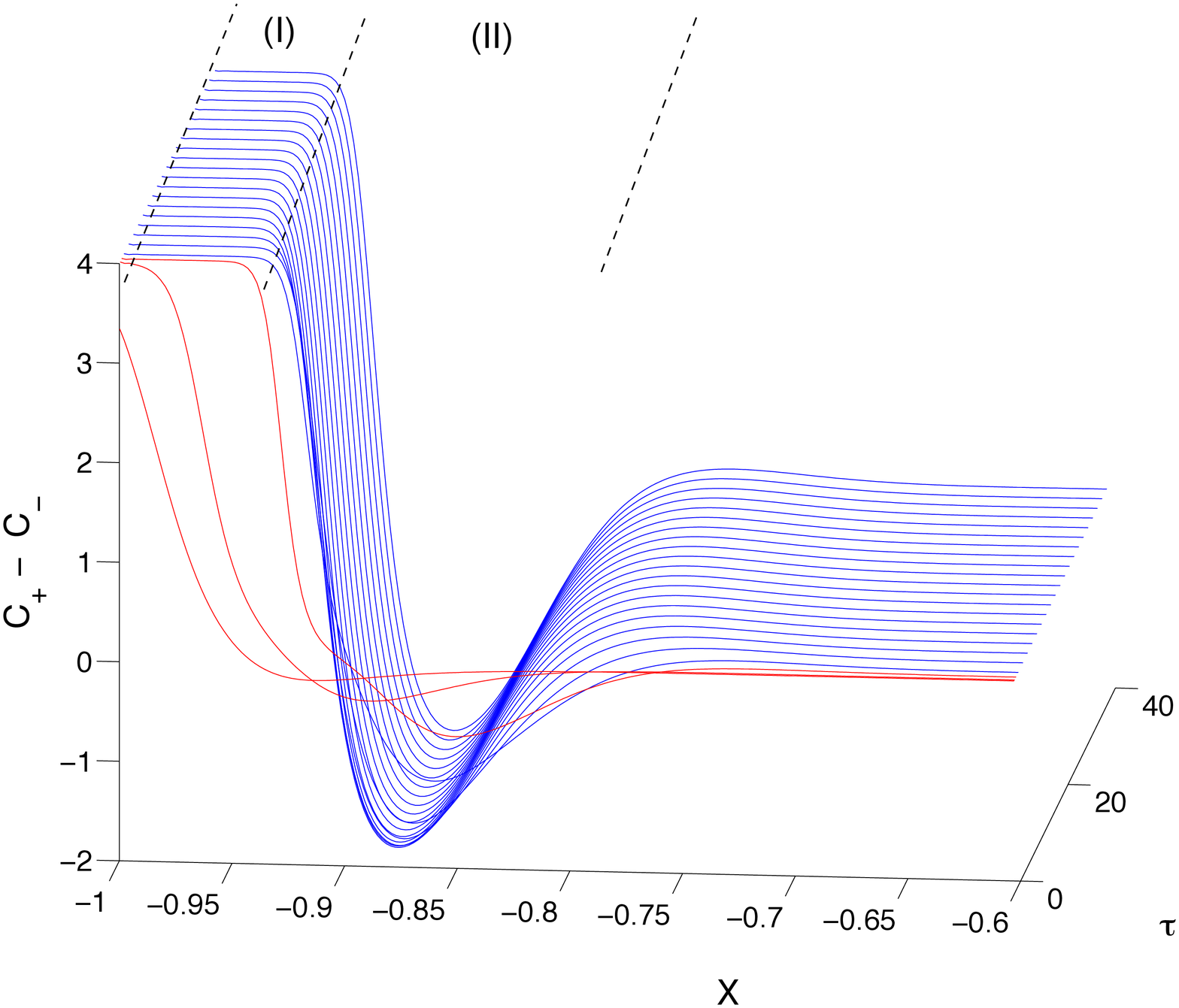}}
\caption{(a) The total density, $C_+ + C_-$, and (b) charge density $C_+ - C_-$, as functions of distance $X=x/L$ from the electrode and time after an applied step voltage $V=40 V_T$ with $V_T=k_BT/e$. Here $\gamma = 0.25$, $l_c/l_D = 10$,  and $L/l_D=100$, red/blue curves denote the first/second charging regimes, and the arrow shows the increasing density deficit in the first charging regime. Numerical solution of Eqs.~(\ref{AKK_overscreening})-(\ref{mod_PNP}) is performed using \texttt{pdepe} in Matlab.}
\label{numerical_sol}
\end{figure}


We turn our attention to a simple problem to gain some insight into the characteristic behaviour of (\ref{AKK_overscreening})-(\ref{mod_PNP}): an ionic liquid with bulk cation/anion concentration $c_0$ bounded by two parallel, blocking electrodes at $x =-L,L$. Initially the concentrations of the two ion species are uniform, and a step voltage of amplitude $2V$ is applied at $t=0^+$. Introducing the Debye length $l_D = 1/\sqrt{8 \pi c_0 l_B}$ and dimensionless packing parameter in the bulk~\cite{kornyshev2007double, kilic2007steric} $\gamma = v c_0$, we introduce the dimensionless variables $\tau = (D/L l_D) t$, $X = x/L$, $C_{\pm} = c_\pm/c_0 $. The no-flux conditions at the electrodes read
\begin{equation}
\left[ \pm C_{\pm} (1- \gamma C) \frac{\partial u}{\partial X} + (1-\gamma C) \frac{\partial C_\pm}{\partial X} + \gamma C_\pm \frac{ \partial C}{\partial X} \right]_{X=\pm 1}=0.  
\end{equation} 
At the electrodes surface, we posit that the classical Gauss law  $\pm \epsilon u_x = 4 \pi \sigma $ holds at $X = \pm 1$, with $\sigma$ the (dimensional) surface charge density, and $\epsilon$ the dielectric constant of the medium \cite{bazant2011double, storey2012effects}. This condition, together with the constant potential condition gives
\begin{equation}
u(X=\pm1,\tau) = \pm V,\; u_{XXX}(X=\pm1,\tau) = 0, \; \tau>0.
\label{bc_V}
\end{equation} 
The initial conditions are 
\begin{equation}
C_\pm(X,0) = 1, \; X\in[-1,1]. 
\end{equation}
To avoid complications of double layer overlap, we consider widely separated electrodes taking $L/l_D = 100$. We take $l_c = a = v^{1/3}$, Bjerrum length $l_B = 50$\AA, ion diameter $a=5$\AA~and $\gamma = 0.25$ (see \emph{e.g.} \cite{bazant2011double} though the qualitative behaviour reported below is not sensitive to $\gamma$); we therefore have $l_c/l_D = \sqrt{8 \pi \gamma l_B /a} \approx 10$. 

Figure \ref{numerical_sol} shows that charging proceeds through two distinct regimes: First, the (negative) electrode attracts cations from the vicinity and expels anions, resulting in a dense, ``compact layer'' of cations near the electrode that overcompensates the surface charge (region I in Figure \ref{numerical_sol}). Ion diffusion is hindered as the mobility matrix (\ref{eq:M}) vanishes in regions of high density. As a result, the total density reaches a minimum away from the compact layer (\emph{c.f.} red arrow in Figure \ref{numerical_sol}a). In the second stage, anions arrive from the bulk to screen the now net-positive compact layer. This flux fills the total density deficit near the compact layer incurred in the first charging regime, creating a region of negative charge density and in fact excess total density (region II in Figure \ref{numerical_sol}b). 

\begin{figure}
\includegraphics[scale=0.23]{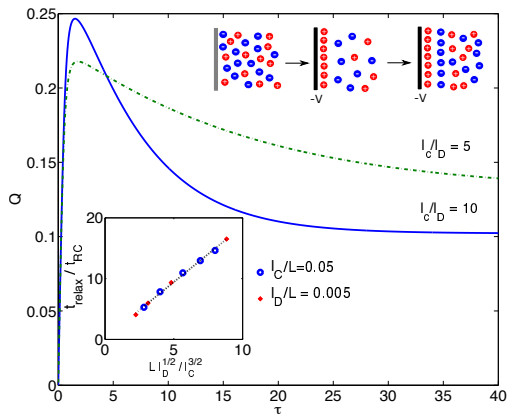}
\caption{(Main figure) The total charge as a function of time for different values of $l_c/l_D$ with $L/l_D = 100$, $\gamma=0.25$ fixed. (Inset) The equilibrium relaxation time $\tau_{\mathrm{relax}}$ as a function of $l_c$ and $l_D$ (obtained by fitting numerical results to an exponential decay). } 
\label{numerical_sol_Q}
\end{figure}

A key measure of practical interest is the integrated total diffuse charge,
\begin{equation}
Q(\tau) = \int_{-1}^{0} \left[C_+(X,\tau)  - C_-(X,\tau) \right]\; \mathrm{d}X. 
\end{equation}
Note that the overall system is electroneutral, therefore the total charge of the ions is equal and opposite to the surface charge. $Q$ is therefore the charge accumulated at the anode, which is equal and opposite in sign to the charge accumulated at the cathode. Figure \ref{numerical_sol_Q} shows that, as charging proceeds, the total charge initially increases, corresponding to the formation of the compact layer. However, arrival of anions in the second charging regime decreases the charge to the final equilibrium value. This charging mechanism is schematically illustrated in the inset of Figure \ref{numerical_sol_Q}. The correlation length $l_c/l_D$ controls the extent of charge oscillation and thus of overcompensation of electrode surface charge by the compact layer. Therefore, decreasing the correlation length reduces the extent of charge overcompensation and also the peak diffuse charge.

Further insights into the charging process can be obtained by noting that the initial rise in charge occurs over $\tau = O(1)$. In dimensional terms this corresponds to $t_{RC} = L l_D/D$, the usual $RC$ time constant \cite{bazant2004diffuse}, corroborating the fact that the peak has its origin in the formation of the diffuse layer. Numerical experimentation (see inset of Figure \ref{numerical_sol_Q}) suggests that the late-stage exponential relaxation of the charge to equilibrium has a distinctly different timescale 
\begin{equation}
\label{eq:trelax}
t_{\mathrm{relax}} = \frac{L^2}{D}\left(\frac{l_D}{l_c}\right)^{3/2}
\end{equation}
This scaling suggests that the decay in the stored charge comes from the formation of charge oscillations: $L^2/D$ gives the decay time due to diffusion of ions through the electrochemical cell, and this is rescaled by $(l_D/l_c)^{3/2}$ where $k_c \sim l_c^{-1}$ is the characteristic wavelength of charge oscillations (\emph{c.f.} Equation (\ref{AKK_overscreening})). 




The non-montonic evolution of $Q(t)$ is in stark contrast to the results predicted by dynamical density functional theory \cite{zhao2011diffuse,yochelis2014transition,yochelis2014spatial}, where the diffuse charge is monotonically increasing. We note that this effect is different from kinetic charge inversion due to double layer overlap \cite{jiang2014kinetic}. The degenerate mobility (\ref{eq:M}) in our approach ensures that the flux due to electrostatic interactions vanishes at close packing, and thus there are distinct regimes of initial charge density polarisation and, at later times, rearrangement of the double layer into cation-rich and anion-rich layers. 

\begin{figure}
\includegraphics[scale=0.23]{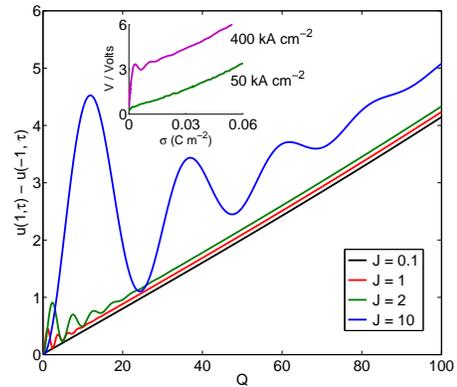}
\caption{The voltage drop across the system evolves non-monotonically under constant current conditions, \eqref{eqn:constcharge}. Here $L/l_D = 100$, $l_c/l_D= 10$ and $\gamma=0.25$, and the total charge $Q = J \tau$. Inset: simulation data from \cite{jiang2014dynamics} for which $\approx 400 \mathrm{~kA cm^{-2}}$ corresponds to dimensionless $J=2$.}
\label{const_curr}
\end{figure}

Qualitatively similar behaviour is obtained under charge-controlled conditions, \textit{i.e.}~imposing a constant current,
\begin{equation}
\frac{\partial u}{\partial X}\Bigg|_{X = \pm 1} = \pm J \tau.
\label{eqn:constcharge}
\end{equation}
Figure~\ref{const_curr} shows that the non-equilibrium double layer rearrangement manifests itself in the non-monotonic evolution of the potential drop across the system when the current density $J$ is large. This qualitatively agrees with recent molecular dynamics simulations \cite{jiang2014dynamics}, but is in contrast to conventional dynamical density functional theory, which again predicts a monotonic increase in potential drop as a function of time.  

In summary, we have derived a continuous model for the dynamics of solvent-free ionic liquids based on coarse-graining a simple exclusion process of interacting particles defined on a lattice. The resulting equations have the structure of a gradient flow with a degenerate mobility function. As examples, these equations were analysed for a system where: (i) a step voltage is applied between widely separated electrodes, and (ii) a constant charging current is applied. Even in these simple cases, our theory differs qualitatively  from previously developed theories for electrolyte solutions. Importantly, we showed that the total diffuse charge is a non-monotonic function of time. Experiments and simulations of the dynamics of ion transport in ionic liquids are currently scarce; we hope that our theory provides a framework to interpret experiments and motivate further investigation. 



\begin{acknowledgments}
We thank S Perkin for discussions about the structure of ionic liquids, and G Oshanin and A A Kornyshev for discussions about kinetic lattice gas systems. This work is supported by an EPSRC Research Studentship to AAL.
\end{acknowledgments}
\bibliography{electrokinetics_ref}

\begin{thebibliography}{44}
\expandafter\ifx\csname natexlab\endcsname\relax\def\natexlab#1{#1}\fi
\expandafter\ifx\csname bibnamefont\endcsname\relax
  \def\bibnamefont#1{#1}\fi
\expandafter\ifx\csname bibfnamefont\endcsname\relax
  \def\bibfnamefont#1{#1}\fi
\expandafter\ifx\csname citenamefont\endcsname\relax
  \def\citenamefont#1{#1}\fi
\expandafter\ifx\csname url\endcsname\relax
  \def\url#1{\texttt{#1}}\fi
\expandafter\ifx\csname urlprefix\endcsname\relax\def\urlprefix{URL }\fi
\providecommand{\bibinfo}[2]{#2}
\providecommand{\eprint}[2][]{\url{#2}}

\bibitem[{\citenamefont{Nemat-Nasser}(2008)}]{nemat2008electrochemomechanics}
\bibinfo{author}{\bibfnamefont{S.}~\bibnamefont{Nemat-Nasser}},
  \bibinfo{journal}{Springer Handbook of Experimental Solid Mechanics} pp.
  \bibinfo{pages}{187--202} (\bibinfo{year}{2008}).

\bibitem[{\citenamefont{Liu et~al.}(2010)\citenamefont{Liu, Liu, Liu, Lin,
  Zhou, Janik, Colby, and Zhang}}]{liu2010influence}
\bibinfo{author}{\bibfnamefont{S.}~\bibnamefont{Liu}},
  \bibinfo{author}{\bibfnamefont{W.}~\bibnamefont{Liu}},
  \bibinfo{author}{\bibfnamefont{Y.}~\bibnamefont{Liu}},
  \bibinfo{author}{\bibfnamefont{J.-H.} \bibnamefont{Lin}},
  \bibinfo{author}{\bibfnamefont{X.}~\bibnamefont{Zhou}},
  \bibinfo{author}{\bibfnamefont{M.~J.} \bibnamefont{Janik}},
  \bibinfo{author}{\bibfnamefont{R.~H.} \bibnamefont{Colby}}, \bibnamefont{and}
  \bibinfo{author}{\bibfnamefont{Q.}~\bibnamefont{Zhang}},
  \bibinfo{journal}{Polymer International} \textbf{\bibinfo{volume}{59}},
  \bibinfo{pages}{321} (\bibinfo{year}{2010}).

\bibitem[{\citenamefont{Lee et~al.}(2013)\citenamefont{Lee, Colby, and
  Kornyshev}}]{colby2013electroactuation}
\bibinfo{author}{\bibfnamefont{A.~A.} \bibnamefont{Lee}},
  \bibinfo{author}{\bibfnamefont{R.~H.} \bibnamefont{Colby}}, \bibnamefont{and}
  \bibinfo{author}{\bibfnamefont{A.~A.} \bibnamefont{Kornyshev}},
  \bibinfo{journal}{Soft Matter} \textbf{\bibinfo{volume}{9}},
  \bibinfo{pages}{3767} (\bibinfo{year}{2013}).

\bibitem[{\citenamefont{Chmiola et~al.}(2006)\citenamefont{Chmiola, Yushin,
  Gogotsi, Portet, Simon, and Taberna}}]{gogotsi:sci:06}
\bibinfo{author}{\bibfnamefont{J.}~\bibnamefont{Chmiola}},
  \bibinfo{author}{\bibfnamefont{G.}~\bibnamefont{Yushin}},
  \bibinfo{author}{\bibfnamefont{Y.}~\bibnamefont{Gogotsi}},
  \bibinfo{author}{\bibfnamefont{C.}~\bibnamefont{Portet}},
  \bibinfo{author}{\bibfnamefont{P.}~\bibnamefont{Simon}}, \bibnamefont{and}
  \bibinfo{author}{\bibfnamefont{P.~L.} \bibnamefont{Taberna}},
  \bibinfo{journal}{Science} \textbf{\bibinfo{volume}{313}},
  \bibinfo{pages}{1760} (\bibinfo{year}{2006}).

\bibitem[{\citenamefont{Largeot et~al.}(2008)\citenamefont{Largeot, Portet,
  Chmiola, Taberna, Gogotsi, and Simon}}]{largeot2008relation}
\bibinfo{author}{\bibfnamefont{C.}~\bibnamefont{Largeot}},
  \bibinfo{author}{\bibfnamefont{C.}~\bibnamefont{Portet}},
  \bibinfo{author}{\bibfnamefont{J.}~\bibnamefont{Chmiola}},
  \bibinfo{author}{\bibfnamefont{P.}~\bibnamefont{Taberna}},
  \bibinfo{author}{\bibfnamefont{Y.}~\bibnamefont{Gogotsi}}, \bibnamefont{and}
  \bibinfo{author}{\bibfnamefont{P.}~\bibnamefont{Simon}},
  \bibinfo{journal}{Journal of the American Chemical Society}
  \textbf{\bibinfo{volume}{130}}, \bibinfo{pages}{2730} (\bibinfo{year}{2008}).

\bibitem[{\citenamefont{Kondrat and Kornyshev}(2011)}]{kondrat:jpcm:11}
\bibinfo{author}{\bibfnamefont{S.}~\bibnamefont{Kondrat}} \bibnamefont{and}
  \bibinfo{author}{\bibfnamefont{A.}~\bibnamefont{Kornyshev}},
  \bibinfo{journal}{J. Phys.: Condens. Matter} \textbf{\bibinfo{volume}{23}},
  \bibinfo{pages}{022201} (\bibinfo{year}{2011}).

\bibitem[{\citenamefont{Simon and Gogotsi}(2013)}]{simon_gogotsi:acr:13}
\bibinfo{author}{\bibfnamefont{P.}~\bibnamefont{Simon}} \bibnamefont{and}
  \bibinfo{author}{\bibfnamefont{Y.}~\bibnamefont{Gogotsi}},
  \bibinfo{journal}{Acc. Chem. Res.} \textbf{\bibinfo{volume}{46}},
  \bibinfo{pages}{1094} (\bibinfo{year}{2013}).

\bibitem[{\citenamefont{Lee et~al.}(2014{\natexlab{a}})\citenamefont{Lee,
  Kondrat, and Kornyshev}}]{kondrat2014single}
\bibinfo{author}{\bibfnamefont{A.~A.} \bibnamefont{Lee}},
  \bibinfo{author}{\bibfnamefont{S.}~\bibnamefont{Kondrat}}, \bibnamefont{and}
  \bibinfo{author}{\bibfnamefont{A.~A.} \bibnamefont{Kornyshev}},
  \bibinfo{journal}{Physical Review Letters} \textbf{\bibinfo{volume}{113}},
  \bibinfo{pages}{048701} (\bibinfo{year}{2014}{\natexlab{a}}).

\bibitem[{\citenamefont{Lee et~al.}(2014{\natexlab{b}})\citenamefont{Lee,
  Vella, Perkin, and Goriely}}]{Lee2014}
\bibinfo{author}{\bibfnamefont{A.~A.} \bibnamefont{Lee}},
  \bibinfo{author}{\bibfnamefont{D.}~\bibnamefont{Vella}},
  \bibinfo{author}{\bibfnamefont{S.}~\bibnamefont{Perkin}}, \bibnamefont{and}
  \bibinfo{author}{\bibfnamefont{A.}~\bibnamefont{Goriely}},
  \bibinfo{journal}{Journal of Physical Chemistry Letters}
  (\bibinfo{year}{2014}{\natexlab{b}}).

\bibitem[{\citenamefont{Gebbie et~al.}(2013)\citenamefont{Gebbie, Valtiner,
  Banquy, Fox, Henderson, and Israelachvili}}]{gebbie2013ionic}
\bibinfo{author}{\bibfnamefont{M.~A.} \bibnamefont{Gebbie}},
  \bibinfo{author}{\bibfnamefont{M.}~\bibnamefont{Valtiner}},
  \bibinfo{author}{\bibfnamefont{X.}~\bibnamefont{Banquy}},
  \bibinfo{author}{\bibfnamefont{E.~T.} \bibnamefont{Fox}},
  \bibinfo{author}{\bibfnamefont{W.~A.} \bibnamefont{Henderson}},
  \bibnamefont{and} \bibinfo{author}{\bibfnamefont{J.~N.}
  \bibnamefont{Israelachvili}}, \bibinfo{journal}{Proceedings of the National
  Academy of Sciences} \textbf{\bibinfo{volume}{110}}, \bibinfo{pages}{9674}
  (\bibinfo{year}{2013}).

\bibitem[{\citenamefont{Fedorov and Kornyshev}(2014)}]{fedorov2014ionic}
\bibinfo{author}{\bibfnamefont{M.~V.} \bibnamefont{Fedorov}} \bibnamefont{and}
  \bibinfo{author}{\bibfnamefont{A.~A.} \bibnamefont{Kornyshev}},
  \bibinfo{journal}{Chemical Reviews} \textbf{\bibinfo{volume}{114}},
  \bibinfo{pages}{2978} (\bibinfo{year}{2014}).

\bibitem[{\citenamefont{Kilic et~al.}(2007)\citenamefont{Kilic, Bazant, and
  Ajdari}}]{kilic2007steric}
\bibinfo{author}{\bibfnamefont{M.~S.} \bibnamefont{Kilic}},
  \bibinfo{author}{\bibfnamefont{M.~Z.} \bibnamefont{Bazant}},
  \bibnamefont{and} \bibinfo{author}{\bibfnamefont{A.}~\bibnamefont{Ajdari}},
  \bibinfo{journal}{Physical Review E} \textbf{\bibinfo{volume}{75}},
  \bibinfo{pages}{021503} (\bibinfo{year}{2007}).

\bibitem[{\citenamefont{Zhao}(2011)}]{zhao2011diffuse}
\bibinfo{author}{\bibfnamefont{H.}~\bibnamefont{Zhao}},
  \bibinfo{journal}{Physical Review E} \textbf{\bibinfo{volume}{84}},
  \bibinfo{pages}{051504} (\bibinfo{year}{2011}).

\bibitem[{\citenamefont{Jiang et~al.}(2014{\natexlab{a}})\citenamefont{Jiang,
  Cao, Jiang, and Wu}}]{jiang2014time}
\bibinfo{author}{\bibfnamefont{J.}~\bibnamefont{Jiang}},
  \bibinfo{author}{\bibfnamefont{D.}~\bibnamefont{Cao}},
  \bibinfo{author}{\bibfnamefont{D.-e.} \bibnamefont{Jiang}}, \bibnamefont{and}
  \bibinfo{author}{\bibfnamefont{J.}~\bibnamefont{Wu}},
  \bibinfo{journal}{Journal of Physics: Condensed Matter}
  \textbf{\bibinfo{volume}{26}}, \bibinfo{pages}{284102}
  (\bibinfo{year}{2014}{\natexlab{a}}).

\bibitem[{\citenamefont{Jiang et~al.}(2014{\natexlab{b}})\citenamefont{Jiang,
  Cao, Jiang, and Wu}}]{jiang2014kinetic}
\bibinfo{author}{\bibfnamefont{J.}~\bibnamefont{Jiang}},
  \bibinfo{author}{\bibfnamefont{D.}~\bibnamefont{Cao}},
  \bibinfo{author}{\bibfnamefont{D.-e.} \bibnamefont{Jiang}}, \bibnamefont{and}
  \bibinfo{author}{\bibfnamefont{J.}~\bibnamefont{Wu}}, \bibinfo{journal}{The
  Journal of Physical Chemistry Letters}  (\bibinfo{year}{2014}{\natexlab{b}}).

\bibitem[{\citenamefont{Yochelis}(2014{\natexlab{a}})}]{yochelis2014transition}
\bibinfo{author}{\bibfnamefont{A.}~\bibnamefont{Yochelis}},
  \bibinfo{journal}{Physical Chemistry Chemical Physics}
  \textbf{\bibinfo{volume}{16}}, \bibinfo{pages}{2836}
  (\bibinfo{year}{2014}{\natexlab{a}}).

\bibitem[{\citenamefont{Yochelis}(2014{\natexlab{b}})}]{yochelis2014spatial}
\bibinfo{author}{\bibfnamefont{A.}~\bibnamefont{Yochelis}},
  \bibinfo{journal}{The Journal of Physical Chemistry C}
  \textbf{\bibinfo{volume}{118}}, \bibinfo{pages}{5716}
  (\bibinfo{year}{2014}{\natexlab{b}}).

\bibitem[{\citenamefont{Marconi and Tarazona}(1999)}]{marconi1999dynamic}
\bibinfo{author}{\bibfnamefont{U.~M.~B.} \bibnamefont{Marconi}}
  \bibnamefont{and} \bibinfo{author}{\bibfnamefont{P.}~\bibnamefont{Tarazona}},
  \bibinfo{journal}{The Journal of Chemical physics}
  \textbf{\bibinfo{volume}{110}}, \bibinfo{pages}{8032} (\bibinfo{year}{1999}).

\bibitem[{\citenamefont{Marconi and Tarazona}(2000)}]{marconi2000dynamic}
\bibinfo{author}{\bibfnamefont{U.~M.~B.} \bibnamefont{Marconi}}
  \bibnamefont{and} \bibinfo{author}{\bibfnamefont{P.}~\bibnamefont{Tarazona}},
  \bibinfo{journal}{Journal of Physics: Condensed Matter}
  \textbf{\bibinfo{volume}{12}}, \bibinfo{pages}{A413} (\bibinfo{year}{2000}).

\bibitem[{\citenamefont{Espa{\~n}ol and
  L{\"o}wen}(2009)}]{espanol2009derivation}
\bibinfo{author}{\bibfnamefont{P.}~\bibnamefont{Espa{\~n}ol}} \bibnamefont{and}
  \bibinfo{author}{\bibfnamefont{H.}~\bibnamefont{L{\"o}wen}},
  \bibinfo{journal}{The Journal of chemical physics}
  \textbf{\bibinfo{volume}{131}}, \bibinfo{pages}{244101}
  (\bibinfo{year}{2009}).

\bibitem[{\citenamefont{Spohn}(1991)}]{spohn1991large}
\bibinfo{author}{\bibfnamefont{H.}~\bibnamefont{Spohn}},
  \emph{\bibinfo{title}{Large scale dynamics of interacting particles}}, vol.
  \bibinfo{volume}{825} (\bibinfo{publisher}{Springer}, \bibinfo{year}{1991}).

\bibitem[{\citenamefont{Bikerman}(1942)}]{bikerman1942xxxix}
\bibinfo{author}{\bibfnamefont{J.}~\bibnamefont{Bikerman}},
  \bibinfo{journal}{The London, Edinburgh, and Dublin Philosophical Magazine
  and Journal of Science} \textbf{\bibinfo{volume}{33}}, \bibinfo{pages}{384}
  (\bibinfo{year}{1942}).

\bibitem[{\citenamefont{Kornyshev}(2007)}]{kornyshev2007double}
\bibinfo{author}{\bibfnamefont{A.~A.} \bibnamefont{Kornyshev}},
  \bibinfo{journal}{The Journal of Physical Chemistry B}
  \textbf{\bibinfo{volume}{111}}, \bibinfo{pages}{5545} (\bibinfo{year}{2007}).

\bibitem[{\citenamefont{Bazant et~al.}(2009)\citenamefont{Bazant, Kilic,
  Storey, and Ajdari}}]{bazant2009towards}
\bibinfo{author}{\bibfnamefont{M.~Z.} \bibnamefont{Bazant}},
  \bibinfo{author}{\bibfnamefont{M.~S.} \bibnamefont{Kilic}},
  \bibinfo{author}{\bibfnamefont{B.~D.} \bibnamefont{Storey}},
  \bibnamefont{and} \bibinfo{author}{\bibfnamefont{A.}~\bibnamefont{Ajdari}},
  \bibinfo{journal}{Advances in colloid and interface science}
  \textbf{\bibinfo{volume}{152}}, \bibinfo{pages}{48} (\bibinfo{year}{2009}).

\bibitem[{\citenamefont{Bazant et~al.}(2011)\citenamefont{Bazant, Storey, and
  Kornyshev}}]{bazant2011double}
\bibinfo{author}{\bibfnamefont{M.~Z.} \bibnamefont{Bazant}},
  \bibinfo{author}{\bibfnamefont{B.~D.} \bibnamefont{Storey}},
  \bibnamefont{and} \bibinfo{author}{\bibfnamefont{A.~A.}
  \bibnamefont{Kornyshev}}, \bibinfo{journal}{Physical Review Letters}
  \textbf{\bibinfo{volume}{106}}, \bibinfo{pages}{046102}
  (\bibinfo{year}{2011}).

\bibitem[{\citenamefont{Gouyet}(1993)}]{gouyet1993atomic}
\bibinfo{author}{\bibfnamefont{J.-F.} \bibnamefont{Gouyet}},
  \bibinfo{journal}{EPL (Europhysics Letters)} \textbf{\bibinfo{volume}{21}},
  \bibinfo{pages}{335} (\bibinfo{year}{1993}).

\bibitem[{\citenamefont{Plapp and Gouyet}(1997)}]{plapp1997surface}
\bibinfo{author}{\bibfnamefont{M.}~\bibnamefont{Plapp}} \bibnamefont{and}
  \bibinfo{author}{\bibfnamefont{J.-F.} \bibnamefont{Gouyet}},
  \bibinfo{journal}{Physical Review Letters} \textbf{\bibinfo{volume}{78}},
  \bibinfo{pages}{4970} (\bibinfo{year}{1997}).

\bibitem[{\citenamefont{Plapp and Gouyet}(1999)}]{plapp1999spinodal}
\bibinfo{author}{\bibfnamefont{M.}~\bibnamefont{Plapp}} \bibnamefont{and}
  \bibinfo{author}{\bibfnamefont{J.-F.} \bibnamefont{Gouyet}},
  \bibinfo{journal}{The European Physical Journal B-Condensed Matter and
  Complex Systems} \textbf{\bibinfo{volume}{9}}, \bibinfo{pages}{267}
  (\bibinfo{year}{1999}).

\bibitem[{\citenamefont{Gouyet et~al.}(2003)\citenamefont{Gouyet, Plapp,
  Dieterich, and Maass}}]{gouyet2003description}
\bibinfo{author}{\bibfnamefont{J.-F.} \bibnamefont{Gouyet}},
  \bibinfo{author}{\bibfnamefont{M.}~\bibnamefont{Plapp}},
  \bibinfo{author}{\bibfnamefont{W.}~\bibnamefont{Dieterich}},
  \bibnamefont{and} \bibinfo{author}{\bibfnamefont{P.}~\bibnamefont{Maass}},
  \bibinfo{journal}{Advances in Physics} \textbf{\bibinfo{volume}{52}},
  \bibinfo{pages}{523} (\bibinfo{year}{2003}).

\bibitem[{\citenamefont{Petrishcheva and
  Abart}(2012)}]{petrishcheva2012exsolution}
\bibinfo{author}{\bibfnamefont{E.}~\bibnamefont{Petrishcheva}}
  \bibnamefont{and} \bibinfo{author}{\bibfnamefont{R.}~\bibnamefont{Abart}},
  \bibinfo{journal}{Acta Materialia} \textbf{\bibinfo{volume}{60}},
  \bibinfo{pages}{5481} (\bibinfo{year}{2012}).

\bibitem[{\citenamefont{Giacomin and Lebowitz}(1996)}]{giacomin1996exact}
\bibinfo{author}{\bibfnamefont{G.}~\bibnamefont{Giacomin}} \bibnamefont{and}
  \bibinfo{author}{\bibfnamefont{J.~L.} \bibnamefont{Lebowitz}},
  \bibinfo{journal}{Physical Review Letters} \textbf{\bibinfo{volume}{76}},
  \bibinfo{pages}{1094} (\bibinfo{year}{1996}).

\bibitem[{\citenamefont{Giacomin and Lebowitz}(1997)}]{giacomin1997phase}
\bibinfo{author}{\bibfnamefont{G.}~\bibnamefont{Giacomin}} \bibnamefont{and}
  \bibinfo{author}{\bibfnamefont{J.~L.} \bibnamefont{Lebowitz}},
  \bibinfo{journal}{Journal of Statistical Physics}
  \textbf{\bibinfo{volume}{87}}, \bibinfo{pages}{37} (\bibinfo{year}{1997}).

\bibitem[{\citenamefont{Giacomin and Lebowitz}(1998)}]{giacomin1998phase}
\bibinfo{author}{\bibfnamefont{G.}~\bibnamefont{Giacomin}} \bibnamefont{and}
  \bibinfo{author}{\bibfnamefont{J.~L.} \bibnamefont{Lebowitz}},
  \bibinfo{journal}{SIAM Journal on Applied Mathematics}
  \textbf{\bibinfo{volume}{58}}, \bibinfo{pages}{1707} (\bibinfo{year}{1998}).

\bibitem[{\citenamefont{Nauman and He}(2001)}]{nauman2001nonlinear}
\bibinfo{author}{\bibfnamefont{E.~B.} \bibnamefont{Nauman}} \bibnamefont{and}
  \bibinfo{author}{\bibfnamefont{D.~Q.} \bibnamefont{He}},
  \bibinfo{journal}{Chemical Engineering Science}
  \textbf{\bibinfo{volume}{56}}, \bibinfo{pages}{1999} (\bibinfo{year}{2001}).

\bibitem[{\citenamefont{Ferguson and
  Bazant}(2012)}]{ferguson2012nonequilibrium}
\bibinfo{author}{\bibfnamefont{T.~R.} \bibnamefont{Ferguson}} \bibnamefont{and}
  \bibinfo{author}{\bibfnamefont{M.~Z.} \bibnamefont{Bazant}},
  \bibinfo{journal}{Journal of The Electrochemical Society}
  \textbf{\bibinfo{volume}{159}}, \bibinfo{pages}{A1967}
  (\bibinfo{year}{2012}).

\bibitem[{\citenamefont{Zeng and Bazant}(2014)}]{zeng2014phase}
\bibinfo{author}{\bibfnamefont{Y.}~\bibnamefont{Zeng}} \bibnamefont{and}
  \bibinfo{author}{\bibfnamefont{M.~Z.} \bibnamefont{Bazant}},
  \bibinfo{journal}{SIAM Journal on Applied Mathematics}
  \textbf{\bibinfo{volume}{74}}, \bibinfo{pages}{980} (\bibinfo{year}{2014}).

\bibitem[{\citenamefont{Chen and Weeks}(2006)}]{chen2006local}
\bibinfo{author}{\bibfnamefont{Y.-G.} \bibnamefont{Chen}} \bibnamefont{and}
  \bibinfo{author}{\bibfnamefont{J.~D.} \bibnamefont{Weeks}},
  \bibinfo{journal}{Proceedings of the National Academy of Sciences}
  \textbf{\bibinfo{volume}{103}}, \bibinfo{pages}{7560} (\bibinfo{year}{2006}).

\bibitem[{\citenamefont{Santangelo}(2006)}]{santangelo2006computing}
\bibinfo{author}{\bibfnamefont{C.~D.} \bibnamefont{Santangelo}},
  \bibinfo{journal}{Physical Review E} \textbf{\bibinfo{volume}{73}},
  \bibinfo{pages}{041512} (\bibinfo{year}{2006}).

\bibitem[{\citenamefont{Bazant}(2013)}]{bazant2013theory}
\bibinfo{author}{\bibfnamefont{M.~Z.} \bibnamefont{Bazant}},
  \bibinfo{journal}{Accounts of chemical research}
  \textbf{\bibinfo{volume}{46}}, \bibinfo{pages}{1144} (\bibinfo{year}{2013}).

\bibitem[{\citenamefont{Storey and Bazant}(2012)}]{storey2012effects}
\bibinfo{author}{\bibfnamefont{B.~D.} \bibnamefont{Storey}} \bibnamefont{and}
  \bibinfo{author}{\bibfnamefont{M.~Z.} \bibnamefont{Bazant}},
  \bibinfo{journal}{Physical Review E} \textbf{\bibinfo{volume}{86}},
  \bibinfo{pages}{056303} (\bibinfo{year}{2012}).

\bibitem[{\citenamefont{Bazant et~al.}(2004)\citenamefont{Bazant, Thornton, and
  Ajdari}}]{bazant2004diffuse}
\bibinfo{author}{\bibfnamefont{M.~Z.} \bibnamefont{Bazant}},
  \bibinfo{author}{\bibfnamefont{K.}~\bibnamefont{Thornton}}, \bibnamefont{and}
  \bibinfo{author}{\bibfnamefont{A.}~\bibnamefont{Ajdari}},
  \bibinfo{journal}{Physical review E} \textbf{\bibinfo{volume}{70}},
  \bibinfo{pages}{021506} (\bibinfo{year}{2004}).

\bibitem[{\citenamefont{Jiang et~al.}(2014{\natexlab{c}})\citenamefont{Jiang,
  Huang, Zhao, Sumpter, and Qiao}}]{jiang2014dynamics}
\bibinfo{author}{\bibfnamefont{X.}~\bibnamefont{Jiang}},
  \bibinfo{author}{\bibfnamefont{J.}~\bibnamefont{Huang}},
  \bibinfo{author}{\bibfnamefont{H.}~\bibnamefont{Zhao}},
  \bibinfo{author}{\bibfnamefont{B.~G.} \bibnamefont{Sumpter}},
  \bibnamefont{and} \bibinfo{author}{\bibfnamefont{R.}~\bibnamefont{Qiao}},
  \bibinfo{journal}{Journal of Physics: Condensed Matter}
  \textbf{\bibinfo{volume}{26}}, \bibinfo{pages}{284109}
  (\bibinfo{year}{2014}{\natexlab{c}}).

\bibitem[{\citenamefont{Canongia~Lopes and
  P{\'a}dua}(2006)}]{canongia2006nanostructural}
\bibinfo{author}{\bibfnamefont{J.~N.} \bibnamefont{Canongia~Lopes}}
  \bibnamefont{and} \bibinfo{author}{\bibfnamefont{A.~A.}
  \bibnamefont{P{\'a}dua}}, \bibinfo{journal}{The Journal of Physical Chemistry
  B} \textbf{\bibinfo{volume}{110}}, \bibinfo{pages}{3330}
  (\bibinfo{year}{2006}).

\bibitem[{\citenamefont{Santos et~al.}(2007)\citenamefont{Santos,
  Canongia~Lopes, Coutinho, Esperan{\c{c}}a, Gomes, Marrucho, and
  Rebelo}}]{santos2007ionic}
\bibinfo{author}{\bibfnamefont{L.~M.} \bibnamefont{Santos}},
  \bibinfo{author}{\bibfnamefont{J.~N.} \bibnamefont{Canongia~Lopes}},
  \bibinfo{author}{\bibfnamefont{J.~A.} \bibnamefont{Coutinho}},
  \bibinfo{author}{\bibfnamefont{J.~M.} \bibnamefont{Esperan{\c{c}}a}},
  \bibinfo{author}{\bibfnamefont{L.~R.} \bibnamefont{Gomes}},
  \bibinfo{author}{\bibfnamefont{I.~M.} \bibnamefont{Marrucho}},
  \bibnamefont{and} \bibinfo{author}{\bibfnamefont{L.~P.}
  \bibnamefont{Rebelo}}, \bibinfo{journal}{Journal of the American Chemical
  Society} \textbf{\bibinfo{volume}{129}}, \bibinfo{pages}{284}
  (\bibinfo{year}{2007}).

\end{thebibliography}

\end{document}